\documentstyle[graphics,epsfig,floats,aps]{revtex}

\begin{document}
\draft

\catcode`\@=11 \catcode`\@=12
\twocolumn[\hsize\textwidth\columnwidth\hsize\csname@twocolumnfalse\endcsname
\title{The Effect of the Tilted Field in the Fractional Quantum Hall
Systems:\\ Numerical Studies for the Solid-liquid Transition}
\author{Yue Yu$^1$ and Shijie Yang$^2$}
\address{$^1$ Institute of Theoretical Physics, Chinese Academy of
Sciences, P.O. Box 2735, Beijing 100080, China}
\address{$^2$ Center for Advanced Study, Tsinghua University,
Beijing 100084, China}

\date{\today}
\maketitle

\begin{abstract}
We construct a generalized Laughlin-liquid wave function and a
variational electron solid wave function when the magnetic field
is tilted. The energy of the liquid state is evaluated by Monte
Carlo methods while the energy of the solid state is calculated by
the optimization. Comparing these two energies for a given tilted
angle $\theta$, it is seen that the critical filling factor
$\nu_c$ of $\theta$ of the solid-liquid transition increases as
the tilted angle. The implication to the experiment is that: i)
the insulating phase may harder be melt as $\nu\to 1/5$ such that
the width of the valley of the longitudinal resistance may become
narrow as the filed is tilted; ii) it is expected that even in the
vicinity of $\nu=1/3$ for the electron system in the presence of
the tilted field, the insulating phase may be observed.
\end{abstract}

]

After the fractional quantum Hall effects (FQHE) \cite{Tsui} was
revealed, the measurement in a tilted field has become an
established technique. Mostly, the tiled field is used to explore
the spin related effects \cite{Cha}. It also affected the
transport of the bilayer system due to the in-plane field couples
to the psuedospin degrees of freedom \cite{Eis}. The high Landau
level stripe phase is strongly influenced by the tilted field
\cite{Eis1}. For the spin polarized system, the tilted field
effects are produced by either the finite thickness of the layer
\cite{Syp} or the subband Laudau level coupling \cite{Cha2}. For
the later case, only a numerical diagonalization for a small
system was carried out. One of focuses in the present work is
trying to construct a Laughlin-type many-particle wave function
when the magnetic field is tilted.

In the early days that Laughlin's wave function was established, a
set of the numerical works to check Laughlin's original wave
function was presented\cite{Laugh,Leav,Morf}. Compared to the
energy of the electron solid state, which is estimated by a
variational method \cite{Lam}, it was found that the critical
filling factor $\nu_c$ is around $\frac{1}{6.5}$. Considering the
Landau level mixing, it was said that the electron solid state may
appear in a much larger filling factor \cite{San,Zhu}. The finite
layer thickness can soften the short range behavior of the Coulomb
interaction and thus leads to favor the competing electron solid
\cite{Sha,San,Pri}. However, at the exact $\nu=1/m$, the
Laughlin-Jastrow correlation for the electron solid may leads to
the melt of the electron solid \cite{Fer} while the energy of the
liquid state at an arbitrary filling factor is higher than that
from the interpolating between the odd-denominator filling factor
\cite{Hal} , which may explain an reentrant insulating phase near
$\nu=1/5$ for the electron \cite{Saj} and $\nu=1/3$ for the hole
\cite{San}.

In this work, we plan to study the competition between the solid
and liquid states as vary of the tilted field through the subband
Landau level coupling. We perform Monte Carlo simulations for a
generalized Laughlin state in the presence of the tilted field. As
expected, the liquid state energy is lowered as increasing the
tilted angle. Meanwhile, we use a variational wave function for
the solid state in the tilted field and optimize the corresponding
energy. The solid state energy is also lowered as increasing the
tilted angle, whose magnitude lowered is always larger than that
of liquid state energy. We can see that the critical filling
factor for the solid-liquid transition may be considerably lifted.
In this work, we do not consider the Laughlin-Jastrow correlation
to the electron solid yet, which leads to an instability of the
electron solid exactly at $\nu=1/m$ for $m$ an odd integer
\cite{Fer}. Thus, although the tilted field effect may not destroy
the liquid state at the exact filling factor $1/5$, the
insulator-quantum Hall liquid transition may become harder to be
melt as $\nu\to 1/5$ such that the width of the valley of the
longitudinal resistance may be narrowed as the tilted angle
increases. Taking the energy lowered by the Laudau level mixing in
the zero tilted field \cite{Zhu} as an estimation, one may even
expect that the insulating phase will be observed near $\nu=1/3$
for the electron systems in the presence of the tilted field.

To consider the subband effect, we assume a quasi-two-dimensional
electron gas confined by a harmonic potential with
$V(z)=\frac{1}{2}m_e\Omega^2z^2$, where $m_e$ is the electron band
mass and $\Omega$ is the  frequency of the harmonic potential in
the perpendicular direction $z$. The applied magnetic field is
tilted by an angle $\theta$ to the two-dimensional $x-y$ plane.
The single-particle problem has been solved in literature
\cite{lit,Yu}. To construct a many-particle Laughlin-type state
for the convenience of the numerical calculation, we take an
alternative way. Working in the symmetric gauge, although the
rotational symmetry of two dimensions is broken by the in-plane
field, there is still a conserved quantity, $L_\xi=b_\xi^\dagger
b_\xi$ no matter whether the magnetic field is tilted, where
$\xi=x+iy$ and
$b_\xi^\dagger=\frac{1}{\sqrt{2}}(-\partial_{\bar\xi}+\frac{1}{2}\xi)$
is the generator of the magnetic translation. The eigen state of
$L_\xi$ thus is the same as that in $B_\parallel=0$. The eigen
state of the single-particle Hamiltonian for $B_\parallel\ne 0$,
then, can be the linear combination of the degenerate states with
a given eigen value  $m_L$ of $L_\xi$,
\begin{eqnarray}
&&\Phi_{m_L}(\vec r)=\sum_{n_1,n_2}C^{\tilde n_1\tilde n_2}
_{n_1n_2}\Psi_{n_1n_2m_L}(\vec r),\\ &&\Psi_{n_1n_2m_L}(\vec
r)=f_{n_1n_2m_L}(\vec r)
e^{-\frac{1}{4}|\xi|^2-\frac{1}{2}{z'}^2}\nonumber\\
&&~~~~~~~~~~~~~~\propto a^{\dagger n_1}_\xi a^{\dagger n_2}_z
b^{\dagger m_L}_\xi e^{-\frac{1}{4}|\xi|^2-\frac{1}{2}{z'}^2},
\nonumber
\end{eqnarray}
where the unit of the magnetic length
$l_B=\sqrt{\frac{c\hbar}{eB_\perp}}=1$ has been used and
$a^\dagger_\xi=\frac{1}{\sqrt
2}(-\partial_\xi+\frac{1}{2}\bar\xi)$ and
$a^\dagger_z=\frac{1}{\sqrt 2}(-\partial _{z'}+z')$ are the
harmonic operators with the frequencies
$\omega_c=\frac{eB_\perp}{m_ec}$ and $\Omega$, respectively;
$z'=\hat \Omega^{1/2} z$ with $\hat \Omega=\tilde\Omega/\omega_c$,
$\tilde\Omega^2=\Omega^2+\omega_\parallel^2$ and
$\omega_\parallel==\frac{eB_\parallel}{m_ec}$. The combination
constants $C^{\tilde n_1\tilde n_2} _{n_1n_2}$ can be determined
by a standard method to find the common eigen states of two
commutated operators. For the ground state, one denotes
$C_{n_1n_2}^{00}=C_{n_1n_2}$, for example, $C_{10}=C_{01}=0$ and
\begin{eqnarray}
C_{11}=-\frac{\omega_c+\tilde\Omega-\omega_+-\omega_-}
{\tilde\omega}C_{00},
\end{eqnarray}
for $C_{00}$ is determined by the normalized condition. Other
$C_{n_1n_2}$ can be deduced by recursive relations. Here
$\omega_\pm^2= \frac{1}{2}(\tilde\Omega^2+\omega_c^2)
\pm\frac{1}{2}\sqrt{(\tilde\Omega^2-\omega_c^2)^2
+4\tilde\omega^2\tilde\Omega\omega_c}$ are the corresponding
frequencies to the diagonalized harmonic operators.

For the liquid state, following Laughlin's track in writing down
his wave function, the required incompressible wave function for
the odd denominator filling factor $\nu=1/m$ has the form
\begin{eqnarray}
\Psi_m(\vec r_1,...,\vec r_2)&=&\prod_i(\sum_{n_1n_2}C_{n_1n_2}
a_{\xi_i}^{\dagger n_1}a_{z_i}^{\dagger n_2})\nonumber\\
&\times&\prod_{i<j}(\xi_i-\xi_j)^m
e^{-\frac{1}{4}\sum_i(|\xi_i|^2+2{z'_i}^2)} \nonumber\\ &\equiv&
\hat P^\dagger(\vec r_1,...,\vec r_N)|m\rangle,
\end{eqnarray}
where $|m\rangle$ is the common Laughlin state. To see the
incompressibilty, one defines the electron density for this state
\begin{eqnarray}
\rho_e(\vec r)=\langle m|\hat P\sum_i\delta^{3}(\vec r-\vec
r_i)\hat P^\dagger| m\rangle/\langle m|\hat P\hat P^\dagger |
m\rangle.
\end{eqnarray}
Since $\langle m|[\hat P,\sum_i\delta^{3}(\vec r-\vec
r_i)]|m\rangle=D_r\langle m|m\rangle=0$ (see (\ref{D})) , it is
easy to show $\rho_e(\vec r)=\langle m|\sum_i\delta^{2}(\vec
r-\vec r_i)| m\rangle/\langle m|m\rangle$, which exactly equals to
the density of the Laughlin state for $B_\parallel=0$. Thus, the
state is incompressible.

In the same spirit, for the solid state, the variational wave
function  reads
\begin{eqnarray}
\Psi_c(\vec r_i)=\hat P^\dagger \psi^c(\xi_i)e^{-\frac{1}{2}\sum_i
{z'_i}^2},
\end{eqnarray}
where $\psi^c$ is the variational wave function\cite{Lam} for
$B_\parallel=0$
\begin{eqnarray}
\psi^c&=&\exp\biggl[\frac{1}{4}\sum_{ij}{u_iB_{ij}u_j}\biggr]\nonumber\\
&\times& \exp\biggl[-\frac{1}{4}\sum_i(|u_i|^2
+(\bar\xi_iR_i+\xi_i\bar R_i)]\biggl],
\end{eqnarray}
where $u_i$ is the site fluctuations of the electrons away from
the equilibrium sites $R_i$ and $B_{ij}$ are the variational
parameters. Similar to Lam and Girvin's, the optimal parameters,
in the momentum space, are given by
\begin{eqnarray}
B_{\vec q}
=e^{i\theta_q}\frac{\omega_L(q)-\omega_T(q)}{\omega_L(q)+\omega_T(q)},
\end{eqnarray}
where $\theta_q$ is a phase factor and  $\omega_L(q)$ and
$\omega_T(q)$ are the effective longitudinal and transverse phone
frequencies, which are determined by the dynamic matrix ${\cal D}$
\cite{Lam,Zhu}. Since the field is tilted, ${\cal D}$ in the
present case is related to an effective potential
\begin{eqnarray}
&&V_{eff}(\vec r-\vec r')={\cal P}_{rr'}V(\vec r-\vec
r')\nonumber\\&&=[1 -\frac{1}{{\cal N}}(D_r^\dagger D_r+
D_{r'}^\dagger D_{r'})+\frac{1}{{\cal N}^2}D_r^\dagger D_r
D_{r'}^\dagger D_{r'}\biggr]V(\vec r-\vec r'),\nonumber\\
&&D_r=\sum_{n_2=0;n_1,l=1}C_{n_1n_2}{z'}_i^{n_2}C^l_{n_1}
(\bar\xi/2)^{n_1-l}\partial_\xi^{n_1-l},\label{D}
\end{eqnarray}
where ${\cal N}=\sum_{n_1n_2} C_{n_1n_2}^2$.

The energies we shall compare for the solid and liquid states are
\begin{eqnarray}
E_{c,m}=\frac{e^2}{\epsilon N}\langle
\Psi_{c,m}|\sum_{i<j}\frac{1}{|\vec r_i-\vec r_j|}|
\Psi_{c,m}\rangle. \label{energy}
\end{eqnarray}
Since the analytic single-particle wave function in the symmetric
gauge is not ready yet, $C_{n_1n_2}$ has to be truncated
\cite{Trun}. Fortunately, we find that for the practical value of
$\Omega/\omega_c$ and a modest tilted angle, the first order
approximation, i.e., considering $C_{11}$-contribution only, may
be a very good one. The expect energy of the approximation single
particle state compares to the exact ground state energy only a
few percentage higher \cite{number}. Thus, we consider the subband
and the tilted field contribution from $C_{11}$ only.

For the liquid state, we evaluate the energy in (\ref{energy}) by
Monte Carlo methods. The detailed simulation technique has been
described by Morf and Halperin \cite{Morf} for the disk geometer.
We consider a finite system. The positive charge neutralizing the
electron is uniformly distributed on a disk of radius $ R_m=
\sqrt{2 m N}l_B$. The energy calculated, at Monte Carlo step $s$,
is
\begin{eqnarray}
E_N^{(s)}=\frac{e^2}{\epsilon N}\sum_{i,j=1}^N\frac{1}{|\vec
r^{(s)}_i-\vec r^{(s)}_j|}+\frac{1}{N}\sum_{i=1} ^N U(\vec
r_i^{(s)}),
\end{eqnarray}
where $\vec r_i^{(s)}$ is the position of the $i$-th electron in
the $s$-th Monte Carlo step; $ U(\vec r_i)$ is the energy of the
$i$-th electron interacting to the neutralizing background:
\begin{eqnarray}
U_B(\vec r_i)=-\frac{e^2}{\epsilon}\int_{r\leq R_m}\frac{\bar
\rho_e}{|\vec r-\vec r_i|},
\end{eqnarray}
where $\bar\rho_e=\frac{\nu}{2\pi l_B^2}$ is the average electron
density. In the Monte Carlo simulation, we have used the unit of
the ion-disk radius $R_0=(\pi\bar\rho_e)^{-1/2}$. The probability
distribution we used in the Monte Carlo simulation has included
the tilted field effect, i.e.,
\begin{eqnarray}
|\Psi_m|^2&=&\biggl|\prod_i(\sum_{n_1n_2}C_{n_1n_2})
a_{\xi_i}^{\dagger n_1}a_{z_i}^{\dagger
n_2}\prod_{i<j}(\xi_i-\xi_j)^m\nonumber\\&\times&
e^{-\frac{1}{4}\sum_i(|\xi_i|^2+2{z'_i}^2)}\biggr|^{2}=e^{-H_m}.
\end{eqnarray}
In the real calculation, the perpendicular component $z'$ can
always integrated out by hands. We identify this energy of the
$s$-th Monte Carlo step as the liquid state energy per electron.
The simulations were done for the finite systems $N=20,30,42$ and
$71$. For $N=20$ and $30$, this energy was computed by 50
independent Monte Carlo simulations, each of which consists of
$170,000$ Monte Carlo steps. For $N=42$ and $71$, the
corresponding numbers are $30$ and $110,000$. The results for
$E_m$ are fitted by the polynomials
\begin{eqnarray}
E_m\approx A+B_1/\sqrt N+B_2 /N,
\end{eqnarray}
where and in the following, the energy unit is $e^2/\epsilon l_B$.
We identify the zero-th order coefficient $A$ as the result of
thermodynamic limit. The values of $A(\nu)$ of $\theta$ for
$\Omega/\omega_c=0.8$ and $0.4$ are listed in Table I.

\vspace{0.3cm}

{\small Table I(a) The thermodynamic limit of $E_m$ for
$\Omega/\omega_c=0.8$.}

\vspace{0.3cm}

\begin{tabular}{ccccc}
\hline\hline $~~\theta~~$&$~~A(1/3)~~$&$~~A(1/5)~~$
&$~~A(1/7)~~$&$~~A(1/9)~~$\\ \hline
0&-0.4099&-0.3275&-0.2809&-0.2499\\ \hline
$\pi/7$&-0.4131&-0.3292&-0.2820&-0.2509\\ \hline
$\pi/5$&-0.4146&-0.3300&-0.2825&-0.2511\\ \hline
$\pi/4$&-0.4163&-0.3307&-0.2830&-0.2514\\ \hline\hline
\end{tabular}

\vspace{0.3cm}

{\small Table I(b) The thermodynamic limit of $E_m$ for
$\Omega/\omega_c=0.4$.}

\vspace{0.3cm}

\begin{tabular}{ccccc}
\hline\hline $~~\theta~~$&$~~A(1/3)~~$&$~~A(1/5)~~$
&$~~A(1/7)~~$&$~~A(1/9)~~$\\ \hline
$\pi/8$&-0.4151&-0.3303&-0.2827&-0.2512\\ \hline
$\pi/5$&-0.4176&-0.3315&-0.2835&-0.2518\\ \hline
$\pi/4$&-0.4258&-0.3356&-0.2862&-0.2537\\ \hline\hline
\end{tabular}

\vspace{0.3cm}

For $\theta=0$, $A$ is independent of $\Omega/\omega_c$, e.g., for
$\nu=1/3$, we have $A=-0.4099$, which is consistent with results
by Levesque et. al \cite{Leav} and Morf and Halperin \cite{Morf}.

According to the results of the thermodynamic limit, one can
figure out the energies dependent on the filling factor $\nu$ and
the tilted angle $\theta$. For $\Omega/\omega_c=0.8$, one can fit
the energy $E_\theta(\nu)$ by the following interpolation formulas
\begin{eqnarray}
&&E_0=-0.782133\nu^{1/2}+0.166\nu^{1.241}-0.009\nu^{2.18},\nonumber\\
&&E_{\frac{\pi}7}=-0.782133\nu^{1/2}+0.154\nu^{1.241}-0.010\nu^{2.18},\nonumber\\
&&E_{\frac{\pi}5}=-0.782133\nu^{1/2}+0.148\nu^{1.241}-0.009\nu^{2.18},\label{fit8}\\
&&E_{\frac{\pi}4}=-0.782133\nu^{1/2}+0.144\nu^{1.241}-0.017\nu^{2.18}.\nonumber
\end{eqnarray}
For $\Omega/\omega_c=0.4$, the fitting results are
\begin{eqnarray}
&&E_{\frac{\pi}8}=-0.782133\nu^{1/2}+0.146\nu^{1.241}-0.010\nu^{2.18},\nonumber\\
&&E_{\frac{\pi}6}=-0.782133\nu^{1/2}+0.138\nu^{1.241}-0.015\nu^{2.18},\label{fit4}\\
&&E_{\frac{\pi}4}=-0.782133\nu^{1/2}+0.110\nu^{1.241}-0.026\nu^{2.18}.\nonumber
\end{eqnarray}
The first one in (\ref{fit8}) is corresponding to $\theta=0$,
which is consistent with the result of Levesque et al \cite{Leav}.
One observes that the exponents in the fit formulas are not
dependent on the tilted field; Neither  are almost the
coefficients of $\nu^{2.18}$ except in the last one of
(\ref{fit4}) which may be because the approximation
single-particle wave function is a little bit poor in this
parameter ($\Delta \varepsilon/\varepsilon\sim 13\%$).

Now, we turn to the electron solid. The strategy we take is the
same as Lam and Girvin's \cite{Lam}: Expanding the Coulomb
interaction to the second order of the fluctuation $u_i$ and
determining the variational parameter $B_{ij}$. Then, calculate
the energy (\ref{energy}) exactly. To the harmonic approximation,
the harmonic term $E_{c,h}$ of the energy is given by
\begin{eqnarray}
E_{c,h}&=&\frac{m}{N}\sum_q(\omega_T(q)+\omega_L(q))^2\\
&=&\frac{m}{N}\sum_q\biggl({\rm Tr}{\cal D}(q)+\sqrt{4{\rm Det}
{\cal D}(q)}\biggr),
\end{eqnarray}
where ${\cal D}$ is the effective dynamic matrix and $\vec q$ is
the wave vector in the Brillouin zone. Comparing to the
uncorrelated energy
$E_{uc,h}=\frac{2m}{N}\sum_q(\omega_T^2(q)+\omega_L^2(q))$, the
fluctuation is lowered the energy because the transverse
 modes are favored while the longitudinal modes are disfavored
\cite{Lam}.

The numerical results of $E_\theta(\nu)$ are as follows. For
$\Omega/\omega_c=0.8$, one has
\begin{eqnarray}
&&E_0=-0.782133\nu^{1/2}+
 0.2430\nu^{3/2}+0.1610\nu^{5/2},\nonumber\\
&&E_{\frac{\pi}7}=-0.782133\nu^{1/2}+ 0.2200\nu^{3/2}+0.1598
\nu^{5/2},\nonumber\\ &&E_{\frac{\pi}5}=-0.782133\nu^{1/2}+ 0.1971
\nu^{3/2}+ 0.1596\nu^{5/2},\label{VE8}\\
&&E_{\frac{\pi}4}=-0.782133\nu^{1/2}+  0.1728  \nu^{3/2}+
0.1594\nu^{5/2}.\nonumber
\end{eqnarray}
For $\Omega/\omega_c=0.4$, it is
\begin{eqnarray}
&&E_{\frac{\pi}8}=-0.782133\nu^{1/2}+ 0.1926 \nu^{3/2}+ 0.1595
\nu^{5/2},\nonumber\\ &&E_{\frac{\pi}6}=-0.782133\nu^{1/2}+
  0.1619 \nu^{3/2}+  0.1593 \nu^{5/2},\label{VE4}\\
&&E_{\frac{\pi}4}=-0.782133\nu^{1/2}+   0.0992\nu^{3/2}+
 0.1592\nu^{5/2},\nonumber
 \end{eqnarray}
The first one in (\ref{VE8}) corresponds to the energy of
$\theta=0$, which is consistent with Lam and Halperin's result
\cite{Lam}. We see that again the coefficients of $\nu^{5/2}$ are
almost not affected by the tilted field.

Using the fitting results (\ref{fit8}) and (\ref{VE8}),
(\ref{fit4}) and (\ref{VE4}), one can obtain the estimated
critical filling factors of $\theta$ (Table II).

\vspace{0.3cm}

{\small Table II(a) The critical filling factors for
$\Omega/\omega_c=0.8$.}

\vspace{0.3cm}

\begin{tabular}{ccc}
 \hline\hline
 $~~~~~~~~~\theta~~~~~~~~~$&$~~~~~~~\Delta E_\theta(1/3)~~~~~~~$&$~~~~~~~~\nu_c~~~~~~~~$\\
 \hline
 0&~-0.015&  6.63\\
 \hline
 $\pi/7$&-0.013 &6.45 \\
 \hline
 $\pi/5$&-0.010&5.58\\
 \hline
 $\pi/4$&-0.007&4.54\\
 \hline\hline
 \end{tabular}

\vspace{0.3cm}

 {\small Table II(b) The critical filling factors for
$\Omega/\omega_c=0.4$.}

\vspace{0.3cm}

\begin{tabular}{ccc}
\hline\hline
 $~~~~~~~~~\theta~~~~~~~~~$&$~~~~~~~\Delta E_\theta(1/3)~~~~~~~$&$~~~~~~~~\nu_c~~~~~~~~$\\
 \hline
 $\pi/8$&-0.009 &5.46 \\
 \hline
 $\pi/6$&-0.007&4.45  \\
 \hline
 $\pi/4$&-0.002&3.46  \\
 \hline\hline
 \end{tabular}

\vspace{0.3cm}

In Table II, $\Delta E_\theta(1/3)$ is the energy difference
between the liquid and solid at $\nu=1/3$, whose magnitude reduces
considerably as the tilted angle increases from 0 to $\pi/4$. In
this work, we did not check the Laughlin-Jastrow correlation to
the electron solid. It has been pointed out that the solid may be
unstable at the exact $\nu=1/5$ due to this correlation
\cite{Fer}. Thus, our variational calculation may not be enough to
completely understand the physics at the exact odd denominator
filling factor $1/m$. Due to the instability, the Hall plateau may
not disappear. However, our calculation indicates that the
insulating phase may harder be melt as $\nu\to 1/5$ such that the
width of the valley of the longitudinal resistance may become
narrow as the tilted angle increases.

On the other hand, although our result did not show the critical
filling factor is shifted to near 1/3, the magnitude of $\Delta
E_\theta(1/3)$ reduces remarkably as the field is tilted.
According to Ref.\cite{Zhu}, after accounting the Laudau level
mixing, the energies of the solid and liquid states for $\theta=0$
are lowered about $0.8\%$ and $0.1\%$ respectively at $\nu\sim
1/3$ and $r_s=\frac{e^2/\epsilon R_0^2}{\nu\hbar\omega_c/2}\sim
2$. The lowering of the energy of the solid is larger than that of
the liquid about $0.002 \frac{e^2}{\epsilon l_B}$, which can
nearly overcome the energy difference $\Delta E_\theta(1/3)$ for
large angles $\theta$ in Table II. If one assumes the energy
lowering due to the Laudau level mixing is independent of the
tilted field and considers the energy of the liquid state for
$\nu$ being nearly but not exactly $1/3$ is in fact higher than
our interpolating estimation \cite{Hal}, it is possible that at a
large tilted angle, the energy of the solid becomes lower than
that of liquid for $\nu\sim 1/3$. Thus it may be anticipated that
in the vicinity of $\nu=1/3$ for the electron system in the
presence of the tilted field, the insulating phase may be observed
for a larger tilted angle. To verify this expected phenomenon,
both further experimental and numerical works are required.

The authors were grateful to the discussions with R. R. Du, Z. B.
Su and F. C. Zhang. This work is partially supported by NSF of
China.


\begin{references}
\bibitem{Trun} Note that a truncated state is still
incompressible with the same density as that of the corresponding
Langhlin state.
\bibitem{number} For examples, for $\Omega/\omega_c=0.8$,  one has
$\Delta \varepsilon/\varepsilon\sim 0.3\%$ for $\theta=\pi/8$ and
$3.6\%$ for $\theta=\pi/4$; For $\Omega/\omega_c=0.4$, it is
$3.1\%$ for $\theta=\pi/7$ and $4.9\%$ for $\theta=\pi/6$. As a
comparison, if one uses the zero tilted field wave function, i.e.,
$C_{11}=0$, one has $\Delta \varepsilon_0/\varepsilon\sim 3.0\%$
for $\theta=\pi/8$ and $11.0\%$ for $\theta=\pi/4$ for
$\Omega/\omega_c=0.8$ ; For $\Omega/\omega_c=0.4$, it is $9.8\%$
for $\theta=\pi/7$ and $12.4\%$ for $\theta=\pi/6$.
\end{references}
\end{document}